\documentclass[aps,prl,twocolumn,superscriptaddress,showpacs,floatfix,longbibliography,nofootinbib]{revtex4-1}
\usepackage[utf8]{inputenc}
\usepackage{graphicx}
\usepackage{color}  
\usepackage{amsmath}
\usepackage{dcolumn}
\usepackage[english]{babel}
\usepackage{amsmath}
\usepackage{amssymb}

\newcolumntype{d}[1]{D{.}{.}{#1}}

\begin{document}

\title{Evidence of Hexadecapole Deformation in Uranium-238\\at the Relativistic Heavy Ion Collider}

\author{Wouter Ryssens}
\affiliation{Institut d’Astronomie et d’Astrophysique, Universit\'e Libre de Bruxelles, Campus de la Plaine CP 226, 1050 Brussels, Belgium}

\author{Giuliano Giacalone}
\affiliation{Institut f\"{u}r Theoretische Physik, Universit\"{a}t Heidelberg, Philosophenweg 16, 69120 Heidelberg, Germany}

\author{Bj\"orn Schenke}
\affiliation{Physics Department, Brookhaven National Laboratory, Upton, NY 11973, USA}

\author{Chun Shen}
\affiliation{Department of Physics and Astronomy, Wayne State University, Detroit, Michigan 48201, USA}
\affiliation{RIKEN BNL Research Center, Brookhaven National Laboratory, Upton, NY 11973, USA}

\begin{abstract}
State-of-the-art hydrodynamic simulations of the quark-gluon plasma (QGP) are unable to reproduce
the elliptic flow of particles observed at the BNL Relativistic Heavy Ion Collider (RHIC) in relativistic
$^{238}$U+$^{238}$U collisions when they rely on information obtained from low-energy experiments for the implementation of deformation in the colliding $^{238}$U ions. We show that this is due to an inappropriate treatment of well-deformed
nuclei in the modeling of the initial conditions of the QGP. Past studies have identified the deformation of the nuclear surface with that of
the nuclear volume, though these are different concepts. In particular, a volume 
quadrupole moment can be generated by both a surface hexadecapole and a surface 
quadrupole moment. This feature was so far neglected in the modeling of heavy-ion collisions, 
and is particularly relevant for nuclei like $^{238}$U, which is both 
quadrupole- and hexadecapole-deformed. With rigorous input from Skyrme density 
functional calculations, we show that correcting for such effects in the 
implementation of nuclear deformations in hydrodynamic simulations 
restores agreement with BNL RHIC data. This brings consistency to the results of
nuclear experiments across energy scales, and demonstrates the impact of the 
hexadecapole deformation of $^{238}$U on high-energy collisions.
\end{abstract}

\maketitle

\paragraph{Introduction.}

The possibility of exploiting the intrinsic deformed shape of atomic nuclei as a means to broaden the scope of ultrarelativistic nuclear collision programs has materialized with the release of data on the collective flow of hadrons 
in collisions of $^{238}$U nuclei (U+U collisions) at the BNL RHIC~\cite{Adamczyk15}.
The quadrupole (ellipsoidal) deformation of this nucleus introduces an elliptical anisotropy in the 
QGP formed in head-on (central) U+U collisions, which enhances
the elliptical modulation (or elliptic flow, $v_2$ \cite{Ollitrault:1992bk,Heinz:2013th}) of the emitted particles in momentum space compared to collisions of less deformed ions, such as $^{197}$Au. 
Effects of this type have been later on identified as well in collisions of other deformed species, namely
 $^{129}$Xe, $^{96}$Ru and $^{96}$Zr \cite{ALICE:2018lao,CMS:2019cyz,ATLAS:2019dct,Abdallah2022,ATLAS:2022dov,Bally:2022vgo}. These observations are of fundamental interest, 
as they allow us to ask whether signatures of the emergent collective properties 
of nuclei can be understood consistently across experimental techniques and energy scales.

To answer this question in general, one should first ensure that 
U+U data is captured by hydrodynamic simulations of the QGP: 
the deformation of $^{238}$U is not only the largest among the 
collided species so far, but it is arguably also the one that is best understood 
by low-energy models and experiments. However, quantitative high-energy 
theory-to-data comparisons have recently led to tensions. Estimates of the elliptic flow resulting from the linear response to an initial QGP eccentricity, $v_2=\kappa_2 \varepsilon_2$, 
show that one obtains an overestimate of U+U data for realistic values of $\kappa_2$ and $\varepsilon_2$~\cite{Giacalone:2020ymy}. 
Large-scale IP-Glasma+MUSIC+UrQMD calculations show good agreement with $v_2$ data 
across energies and collision species, with the exception of central U+U collisions: 
again the predicted $v_2$ overshoots the measurements~\cite{Schenke19,Schenke20}. 
The issue is corroborated by the model-independent analysis of Ref.~\cite{Giacalone21}, arguing that the impact of nuclear deformation can be assessed by comparing 
mean squared $v_2$ coefficients between collision systems. This ratio can be expressed
as:
\begin{equation}
\label{eq:ratio}
r_{\rm Au,U}\{2\}^{2} \equiv \frac{ \langle v_2^2 \rangle_{\rm U+U}}{\langle v_2^2 \rangle_{\rm Au+Au}} = \frac{1 + a_0 (\beta_{2, \rm U}^{\rm WS})^2}{a_1 + a_2 a_0 (\beta^{\rm WS}_{2, \rm Au})^2}, 
\end{equation}
where $\beta^{\rm WS}_{2,\rm Au}$ and $\beta^{\rm WS}_{2,\rm U}$ are Woods-Saxon (WS) deformation parameters, 
defined more precisely below, that reflect the 
quadrupole deformation of both species, while the coefficients $a_{0,1,2}$ represent robust features of the hydrodynamic description of the QGP~\cite{Giacalone21}. 
Setting the parameter $\beta^{\rm WS}_{2}$ to be equal to the quadrupole deformation reported in low-energy literature, one obtains $r_{\rm Au,U}\{2\}^{2} =1.78\pm0.15$, much larger than $1.49\pm0.05$,
the value measured by the STAR Collaboration for head-on (0-1\% central) collisions~\cite{Adamczyk15}.

However, reference~\cite{Giacalone21} along with \textit{all} 
past theoretical studies of high-energy U+U collisions assumes that $\beta^{\rm WS}_{2, \rm U}$ in Eq.~(\ref{eq:ratio}) can be taken from low-energy spectroscopic data, although the deformation extracted from low-energy experiments and the deformation parameter used in the hydrodynamic models are 
different quantities~\cite{Shou15}. Here, we demonstrate that the magnitude of this difference becomes important
for well-deformed nuclei with a significant hexadecapole moment, such as $^{238}$U~\cite{Bemis73,Zumbro84}. 
We discuss the difference between these two concepts of nuclear deformation,
and show that the presence of a hexadecapole moment modifies the appropriate
input for hydrodynamic simulations of the QGP.
We compute, then, realistic nucleon densities via state-of-the-art nuclear
energy density functional (EDF) theory that 
are consistent with low-energy experimental information, to show that an appropriate deformation parameter is $\beta^{\rm WS}_{2,\rm U} \approx 0.25$, 
significantly lower than implemented in previous hydrodynamic studies. 
Through new state-of-the-art simulations of U+U and Au+Au collisions, we 
resolve the tension between high-energy observations and low-energy 
expectations, demonstrating for the first time the impact of the hexadecapole deformation of 
a nucleus on high-energy data.

\paragraph{A tale of two deformations.}

Much of our understanding of the low-energy structure of nuclei hinges on
the notion of deformation: the nuclear density in the intrinsic frame can take 
a variety of shapes. These are typically characterized via dimensionless 
multipole moments of the nucleon density $\rho(\mathbf{r})$:
\begin{align}
\beta_{\ell m} = \frac{4 \pi }{3 R_0^{\ell} A} \int d^3r \, \rho(\mathbf{r}) r^{\ell} Y_{\ell m}(\theta, \phi)\, , \hspace{20pt} \ell\geq2\, ,
\label{eq:betalm}
\end{align}
where $R_0 = 1.2 A^{1/3}$ fm and 
$Y_{\ell m}(\theta, \phi)$ is a spherical harmonic. We also  
define the total deformation at order $\ell$: $\beta^2_{\ell} = \sum_{m=-\ell}^{+\ell} \beta_{\ell m}^2 \, .$
%
%
While the intrinsic body-frame multipole moments $\beta_{\ell m}$ are not directly 
observable, the integral in Eq.~\eqref{eq:betalm} is the expectation value of a 
multipole operator whose matrix elements determine the electromagnetic transition
rates between nuclear levels~\cite{RingSchuck}. Under strict assumptions \cite{Bohr69},
the deformation of an even-even nucleus can be inferred from ground state (g.s.) electric transition rates, $B(E\ell)$:
\begin{align}
\beta_{\ell} = \frac{4 \pi}{(2\ell+1)ZR_0^{\ell}} \sqrt{\frac{B(E \ell)}{e^2}} \, .
\label{eq:BEtransition}
\end{align}
Quadrupole deformation ($\ell=2$) is dominant for essentially all nuclei. 
Octupole ($\ell=3$) and hexadecapole ($\ell=4$) deformations play a role in 
several regions of the nuclear chart~\cite{Nazarewicz81,Butler16,Scamps21}, but measurements 
of $\ell > 2 $ transition rates are scarce.

The nucleus $^{238}$U is the archetype of a well-deformed nucleus 
for which Eq.~\eqref{eq:BEtransition} holds~\cite{Grosse81}.
The recommended value for the $\ell=2$ transition is $B(E2) = 12.19 \pm 0.62 $ e$^2$b$^2$~\cite{Pritychenko16}, corresponding
to $\beta_{2, \rm U} = 0.287 \pm  0.007$. No direct measurements of B(E4) are 
available to date, but several more model-dependent analyses report hexadecapole
deformations ranging between 0.1 and 0.2~\cite{Moss71,Hendrie73}.
We consider the most direct information available to be that of Refs.~\cite{Bemis73} and \cite{Zumbro84}, 
which report $\beta_4 \sim 0.124 \pm 0.033$ and $\beta_4 \sim 0.144 \pm 0.007$
based on Coulomb excitation (Coulex) and muonic x-rays, respectively.
As we will see, theoretical calculations faithfully reproduce the quadrupole 
deformation of $^{238}$U, but favor somewhat larger values of $\beta_{4}$.

The multipole moments of the odd-Z $^{197}$Au cannot be determined from B(E$\ell$) 
values. Instead, Ref.~\cite{Giacalone21} proposed a conservative estimate, 
$\beta_{2, \rm Au} \in [0.1, 0.14]$, based on the predictions of 
various models and the deformations of neighboring species. 
Although not often discussed, models typically predict a non-zero
hexadecapole deformation~\cite{Goriely09,Scamps21} for $^{197}$Au. For instance, a 
recent state-of-the-art multi-reference (MR) EDF calculation finds a triaxial 
g.s.~with $\beta_{2, \rm Au}=0.13$ and $\beta_{4,\rm Au}=0.056$~\cite{Bally23,Benjaminpriv}.

Now, hydrodynamic simulations of high-energy collisions require nuclear densities to model the colliding 
ions. Almost without exception, a Woods-Saxon (WS) parametrization is 
used~\cite{Woods54}:
\begin{align}
\rho^{\rm WS}(\mathbf{r}) = \frac{\rho_0}{1 + \exp\left([r - R(\theta,\phi)]/a\right)} \, ,
\label{eq:WSdensity}
\end{align}
where $\rho_0$ fixes the normalization, $a$ is the surface diffuseness, and the 
angle-dependent radius reads:
\begin{align}
R(\theta,\phi) = R_d \left[ 1 + \sum_{\ell = 2}^{\ell_{\rm max}} \sum_{m = -\ell}^{\ell}  \beta^{\rm WS}_{\ell m} Y_{\ell m}(\theta,\phi)\right]
\label{eq:betaWS}
\end{align}
where $R_d$ is the half-width radius, and $\beta^{\rm WS}_{\ell m}$ are shape parameters
for which we also define a total $\beta_{\ell}^{\rm WS}$.
What has not been fully appreciated so far is that the multipole moments 
$\beta_{\ell m}$ of a WS density are not equal to the values of the 
$\beta^{\rm WS}_{\ell m}$ used to generate them.
The former are linked to expectation values of operators and represent the entire 
nuclear \textit{volume},  while the latter describe
the deformation of the nuclear \textit{surface}. Though tedious, it is possible to express the 
multipole moments of a WS density as a combined power series in the parameters 
$\beta^{\rm WS}_{\ell m}$ and $a/R_d$.
As an example, we give here the 
expression for the quadrupole moment of a density with a
sharp profile ($a = 0$) for which only 
$\beta^{\rm WS}_{20}$ and $\beta^{\rm WS}_{40}$ do not vanish:
\begin{alignat}{3}
\beta_{20} = \frac{R_d^{2}}{R_0^2}  \bigg[
         \beta^{\rm WS}_{20}  
                +  \frac{2}{7} \sqrt{\frac{5}{\pi}}   (\beta_{20}^{\rm WS})^2 
                +  \frac{12}{7\sqrt{\pi}}             \beta_{20}^{\rm WS} \beta_{40}^{\rm WS} 
\bigg]
\, ,
\label{eq:transformation_b2}
\end{alignat}
which is valid up to second order in $\beta^{\rm WS}_{20}$ and to first order in $\beta^{\rm WS}_{40}$, and similar to other equations for liquid-drop-type densities available in the literature~\cite{Hasse88,Ryssens19}.
Equation~\eqref{eq:transformation_b2} shows 
that, if $\beta^{\rm WS}_{20}$ is large, even a small $\beta^{\rm WS}_{40}$ 
will enhance the mismatch between $\beta_{20}$ and $\beta_{20}^{\rm WS}$. 
To our knowledge, this subtlety has never been considered in the modeling of $^{238}$U nuclei in hydrodynamic simulations of the QGP, 
although, as we will show, it impacts significantly the predicted $v_2$ in U+U collisions.
Considering more exotic shapes, with e.g.~finite octupole 
or triaxial deformation, will lead to additional terms 
in Eq.~\eqref{eq:transformation_b2}. The construction of WS densities 
with pre-determined multipole moments is, therefore, a nontrivial task.  

\paragraph{Skyrme-HFB calculations.}

To find WS parameters that better reflect our knowledge of the 
structure of $^{197}$Au and $^{238}$U, we perform Hartree-Fock-Bogoliubov
(HFB) calculations based on an EDF. We limit 
ourselves here to an EDF of the widely used Skyrme type~\cite{Bender03}, 
but report on the predictions of 21 different parametrizations to gauge 
the model spread. These parametrizations come in five families: 
(i) BSkG1/2~\cite{Scamps21,Ryssens22},
(ii) SLy4/5/6~\cite{Chabanat98},
(iii) UNEDF0/1/2~\cite{Kortelainen10,Kortelainen12,Kortelainen14}, 
(iv) SV-min/bas/07/08/10~\cite{Klupfel09},
(v) and SLy5s1-8~\cite{Jodon16}. Together, they are fairly representative of 
the literature.

\begin{figure}[t]
    \centering
    \includegraphics[width=\linewidth]{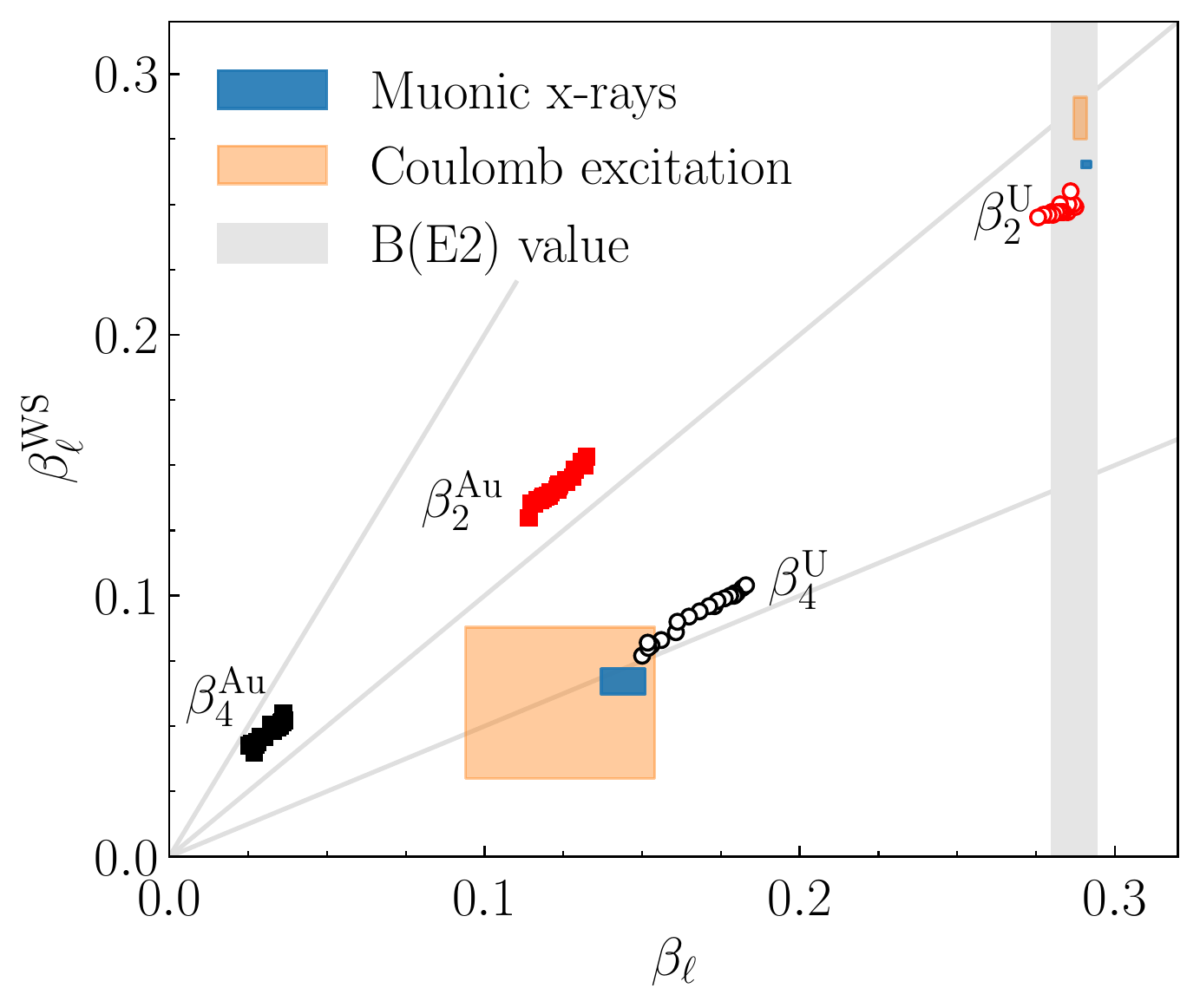}
    \caption{ Total best-fit WS deformation parameters 
              $\beta^{\rm WS}_{\ell}$ versus total deformation $\beta_{\ell}$
              of the mean-field densities obtained with 21 Skyrme 
              parametrizations for $^{197}$Au (full squares) and $^{238}$U 
              (empty circles), for $\ell = 2$ (red symbols) and
              $\ell = 4$ (black symbols). The faint grey lines indicate 
              $\beta^{\rm WS}_{\ell} = \alpha \beta_{\ell}$ for $\alpha =0.5, 1$
              and $2$.
            }
    \label{fig:1}
\end{figure}  

We solve the self-consistent Skyrme-HFB equations for each nucleus and each 
parametrization, relying on a three-dimensional numerical representation of the 
single-particle wavefunctions in coordinate space~\cite{Ryssens16}. The solution
is a many-body Bogoliubov state that minimizes the total energy, whose one-body 
density we use to calculate all multipole moments, $\beta_{\ell m}$, of the 
nuclear ground state. We impose a few symmetry restrictions on the nuclear
shape (see supplement), which in practice determines that 
non-vanishing multipole moments have $\ell$ and $m$ both even. 
Then we obtain deformation parameters $\beta_{\ell m}^{\rm WS}$ by fitting
Eq.~\eqref{eq:WSdensity} to the three-dimensional HFB density \cite{Bally22}. The 
results are shown in Fig.~\ref{fig:1}, displaying the total quadrupole and 
hexadecapole parameters $\beta^{\rm WS}_{\ell=2/4}$ as a function of the 
corresponding total multipole moments for both $^{197}$Au and $^{238}$U. 

We see that the spread in predictions is modest, meaning these 21 Skyrme
parametrizations yield a consistent picture of the structure of these nuclei. 
As expected, the predicted $\beta_{2, \rm U}$ agrees well with that deduced
from the B(E2) value, also shown in the figure as a gray band.
The predicted $\beta_{4,\rm U}$ have a somewhat larger theoretical spread and are in mild tension with the model-dependent experimental information of Refs.~\cite{Bemis73,Zumbro84}, 
which also report values of $\beta_\ell^{\rm WS}$. Overall, the values of 
$\beta_\ell$ correlate linearly with the values of $\beta_{\ell}^{\rm WS}$, 
though with slopes differing from unity. This leads to our central result:  
we see that the values of $\beta_\ell^{\rm WS}$ are consistently lower than the 
values of $\beta_\ell$ for $^{238}$U, due to the contribution of the 
volume hexadecapole deformation to the surface quadrupole deformation.
Indeed, we recover $\beta_{2}^{\rm WS}\approx\beta_{2}\approx0.29$ if we
constrain the EDF calculations to $\beta_4=0$ and values of $\beta_2$ that are 
compatible with low-energy experiment. Our conclusion is that a realistic WS parametrization of 
the g.s. density of $^{238}$U should have $\beta_{2,\rm U}^{\rm WS}\approx 0.25$. 
This value is significantly smaller than the 
volume quadrupole deformation, $\beta_{2, \rm U}$, of this nucleus, and 
 all values of $\beta_{2,\rm U}^{\rm WS}$ used so far in hydrodynamic 
calculations. The difference is a direct consequence of the sizeable hexadecapole
moment of $^{238}$U. In what follows, we demonstrate its impact on the 
interpretation of high-energy data.

For $^{197}$Au, we find a triaxial shape for all parametrizations, with 
$\gamma^{\rm WS}_{\rm Au} = {\rm atan} \left( \sqrt{2}\beta^{\rm WS}_{22, \rm Au}/ \beta^{\rm WS}_{20, \rm Au}\right)\approx 47^{\circ}$, 
agreeing with the recent MR-EDF calculation~\cite{Bally23}. The fitted
WS parameters are larger than the corresponding multipole moments and can
serve as an illustration that the interplay between different deformation modes is indeed
nontrivial.


\paragraph{Understanding RHIC data.}

In what follows, we restrict ourselves to the WS parameters obtained with the 
BSkG2 parametrization; their values, as well as those predicted by the other 
parametrizations are included in the supplementary material. 
We now show that our analysis improves the description of elliptic flow data in U+U collisions. 
We first go back to Eq.~(\ref{eq:ratio}). Combining $a_0=25.6\pm 5$, $a_1 = 1.18 \pm 0.05$ and 
$a_2=1.00^{+0.00}_{-0.05}$ deduced in Ref.~\cite{Giacalone21} with $\beta_{2, \rm Au}^{\rm WS}=0.14$ and our newly-derived $\beta_{2, \rm U}^{\rm WS}=0.25$ leads to $r_{\rm Au,U}\{2\}=1.55 \pm 0.10$, which is finally compatible with the value measured by the STAR Collaboration in U+U collisions at 0-1\% centrality, $1.49 \pm 0.05$ \cite{Adamczyk15}, restoring consistency between high- and low-energy nuclear phenomenology. 

\begin{figure}[t]
    \centering
    \includegraphics[width=\linewidth]{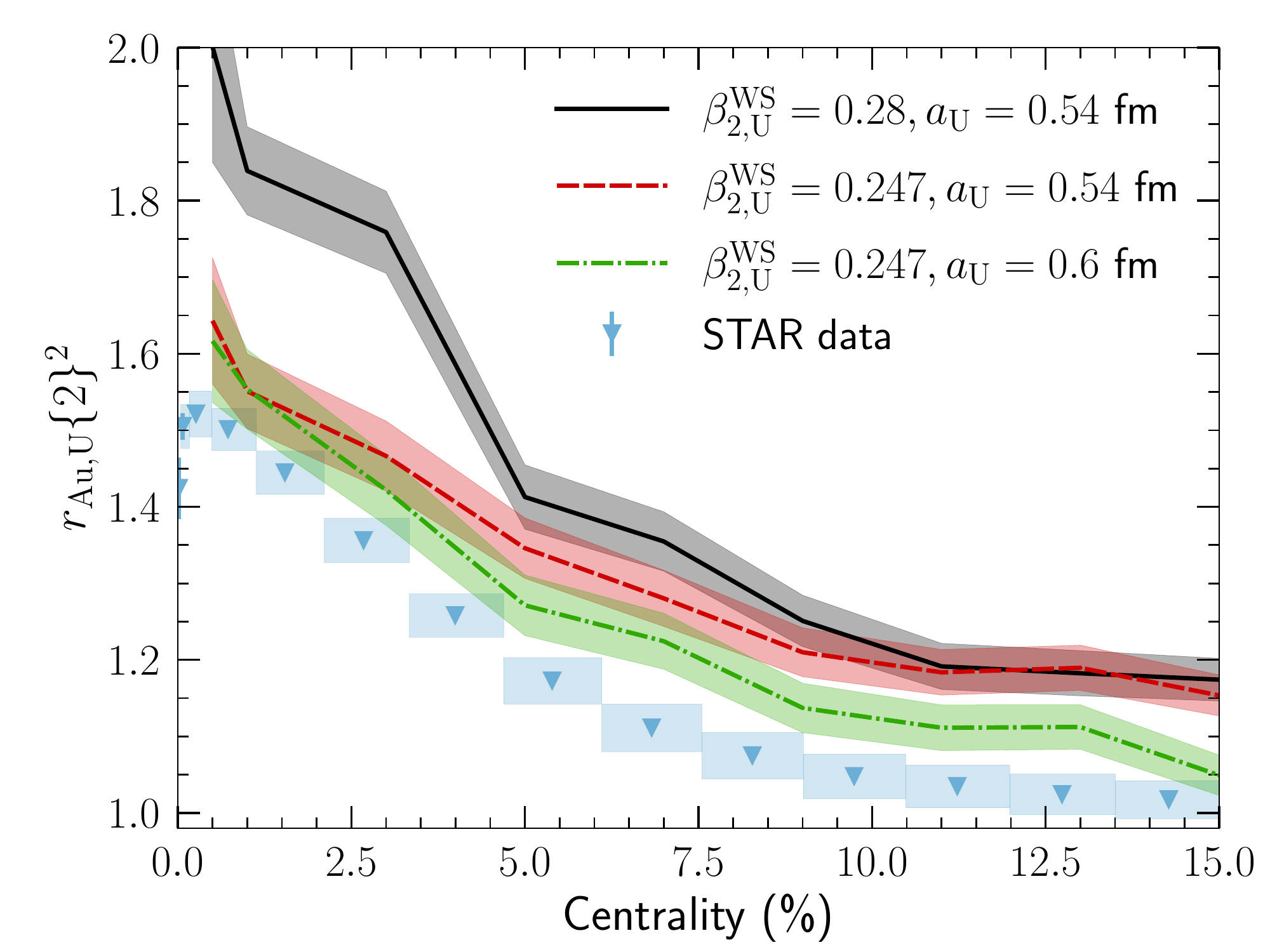}
    \caption{ Ratio of mean squared elliptic flow coefficients, $v_2\{2\}^2 \equiv \langle v_2^2 \rangle$, taken between central U+U and central Au+Au collisions. Symbols: STAR data. Lines are the results of IP+Glasma+MUSIC+UrQMD simulations. Different line styles imply different WS parametrizations for the collided nuclei. Solid line: Original WS parametrizations from Ref.~\cite{Schenke19,Schenke20}. Dashed line: Parametrizations from the Skyrme-EDF calculations of this manuscript. Dot-dashed line: Same as the dashed line but with $a_{\rm U} = 0.60$ fm.}
    \label{fig:3}
\end{figure}

We demonstrate this as well in a direct model application by repeating the IP-Glasma+MUSIC+UrQMD calculations of Ref.~\cite{Schenke20}, this time using the WS parametrizations of $^{238}$U and $^{197}$Au 
from the BSkG2 results (including, in addition, hard-core repulsion among nucleons \cite{footnote}).  We show the predicted  $r_{\rm Au,U}\{2\}^{2}$ 
in Fig.~\ref{fig:3} as a function of collision centrality (dashed line), which is also compared to the original predictions using
$\beta_{2,\rm U}^{\rm WS}=0.28$ (solid line). For 0-1\% collisions, we find that a proper implementation of the deformation of $^{238}$U, 
obtained consistently from state-of-the-art EDF calculations, 
leads to results that are in agreement with both STAR data and the model-independent estimate given in Eq.\,(\ref{eq:ratio_v2}). We have checked in addition that an initial state estimator of the shown ratio, based on eccentricities, exhibits the same behavior.

Moving away from the most central bin, we see that the description of STAR data worsens significantly. This is unlikely to be caused by the deformation parameters, but rather by an inappropriate implementation of the skin of $^{238}$U in the simple WS parametrization. To show this, we repeat our calculation with a 10\% larger parameter $a$ for $^{238}$U (dot-dashed line in Fig.~\ref{fig:3}). This mild correction impacts significantly the centrality dependence of $r_{\rm Au,U}\{2\}^{2}$, without affecting the 0-1\% bin, corroborating the robustness of our main conclusion. 

In the future, one should move away from WS densities and directly input the results of EDF calculations in hydrodynamic simulations, taking into account the difference between spatial distributions of protons and neutrons. For $^{238}$U, this should be especially important due to the strong polarization of its neutron skin across the surface \cite{Liu:2023qeq,Liu:2023rap}, whose effect in high-energy collisions may be similar to an overall broadening of the skin thickness. A recent STAR analysis of the structure of $^{238}$U from ultra-peripheral collisions suggests, for instance, a larger skin than reported in common WS parametrizations \cite{STAR:2022wfe}.

Before concluding, we note that the result shown in Fig.~\ref{fig:3} would remain unchanged if one set 
$\beta_{4,\rm U}^{\rm WS}=0$ in the hydrodynamic simulations, as this shape parameter does not modify the eccentricity fluctuation of the QGP in 
central collisions~\cite{Jia:2021tzt}. A recent transport calculation \cite{Magdy:2022cvt} suggests however
that a modest $\beta_{4, \rm U}^{\rm WS}\approx 0.1$ would impact the quadrangular 
flow, $v_4$, in particular, the so-called \textit{linear}
component of this coefficient in the limit of central collisions. We 
recommend experimental investigations of $v_4$ at high multiplicities and with 
a fine centrality binning as a potential means to independently gauge the magnitude of 
$\beta_{4, \rm U}^{\rm WS}$ at high-energy colliders.

\paragraph{Summary \& outlook.}

The difference between the deformation parameters of a WS density 
($\beta^{\rm WS}_{\ell m}$) and its multipole moments ($\beta_{\ell m}$) is 
particularly large when the nucleus exhibits coexisting deformation modes.
 $^{238}$U represents such a system. Due to its sizeable hexadecapole moment, the appropriate surface deformation 
 parameter, $\beta_{2,\rm U}^{\rm WS} \approx 0.25$, as predicted by state-of-the-art EDF calculations with 21 different Skyrme parametrizations is significantly different
 from the volume deformation, $\beta_{2,\rm U}  \approx 0.28$. Past studies of relativistic U+U collisions 
have not accounted for this subtlety, leading to inconsistencies between BNL RHIC 
data and hydrodynamic calculations in central collisions. Our new simulations demonstrate that our findings resolve these issues:
\begin{align}
\label{eq:ratio_v2}
r_{\rm Au,U}\{2\}^{2,~\rm Ref.~[13]} &= 1.55 \pm 0.10 \,, \\
r_{\rm Au,U}\{2\}^{2,~\rm STAR~data} &= 1.49 \pm 0.05 \,, \\
r_{\rm Au,U}\{2\}^{2,~\rm IP-Glasma~(\beta^{\rm WS}_{2, \rm U}=0.25)} &= 1.63 \pm 0.06 \, .
\end{align}
This is a major 
step towards establishing the consistency of theoretical and experimental 
results across vastly different energy scales. 
The preference of BNL RHIC data for values of $\beta_{2, \rm U}^{\rm WS}$ significantly smaller than reported in spectroscopic data tables provides evidence of the sizeable 
hexadecapole deformation in $^{238}$U, whose phenomenological consequences in high-energy collisions we have reported here for the first time.

That said, there is also some minor tension that could be addressed by the 
low-energy community: essentially all Skyrme parametrizations favor $\beta_{4,\rm U}$
values that are somewhat larger than those determined from muonic x-ray and Coulex 
experiments, which result from model-dependent analyses. We 
hope that the present study will motivate future investigations of the 
hexadecapole moment and the $B(E4)$ g.s. transition rate of $^{238}$U.

Both $^{238}$U and $^{197}$Au are well described by a single
mean-field configuration with a well-defined shape that is reasonably consistent
across models. This is not the case for all other species collided so far: for example
so-called \emph{isobar collisions} at BNL RHIC \cite{Abdallah2022} involve the transitional 
isotopes $^{96}$Ru and $^{96}$Zr, for which a more advanced many-body 
treatment is required, whether based on an EDF or
in an \textit{ab initio} setup.
The analysis of such collisions leads to WS 
shapes that combine sizeable quadrupole and octupole \emph{surface} 
deformation parameters~\cite{Zhang22}. In any effort from the community to confront these data with state-of-the-art calculations, 
corrections due to the interplay between all relevant deformation modes should be accounted for.

As anticipated, one way to achieve this is moving away from simple shape 
parametrizations to generate the initial conditions for hydrodynamic simulations, 
sampling instead nucleon distributions directly provided by nuclear theory. Unfortunately, 
this does not reduce the model dependency of such analysis: predictions for the 
shape of nuclei may vary widely across calculations. A truly model-independent way to construct initial conditions for 
hydrodynamic simulations based on experimental information on nuclear multipole 
moments seems impossible: the $\beta_{\ell m}$ are not coefficients in a
series expansion and do not uniquely characterize the nuclear density.

\paragraph{Acknowledgments.} 
We thank the members of the EMMI Rapid Reaction Task Force 
\textit{``Nuclear physics confronts relativistic collisions of isobars"} for 
valuable input and discussions. G.G. and W.R. acknowledge in particular Benjamin Bally and 
Michael Bender for useful discussions.
The present research benefited from computational resources made available on 
the Tier-1 supercomputer of the Fédération Wallonie-Bruxelles, infrastructure 
funded by the Walloon Region under the grant agreement nr 1117545.
Additional computational resources have been provided by the Consortium des 
Équipements de Calcul Intensif (CÉCI), funded by the Fonds de la Recherche 
Scientifique de Belgique (F.R.S.-FNRS) under Grant No. 2.5020.11 and by the 
Walloon Region.
The heavy-ion simulations used computing resources provided by the Open Science Grid (OSG), supported by the National Science Foundation award \#2030508.
G.G. is funded by the Deutsche Forschungsgemeinschaft 
(DFG, German Research Foundation) under Germany's Excellence Strategy 
EXC2181/1-390900948 (the Heidelberg STRUCTURES Excellence Cluster), within the 
Collaborative Research Center SFB1225 (ISOQUANT, Project-ID 273811115).
This work benefitted from support by the Fonds de la Recherche Scientifique - FNRS and the
Fonds Wetenschappelijk Onderzoek - Vlaanderen (FWO) under the EOS Project No O000422 and O022818F. " 
W.R. gratefully acknowledges financial support from the F.R.S.-FNRS (Belgium).
B.P.S. and C.S. are supported by the U.S. Department of Energy, Office of Science, Office of Nuclear Physics, under DOE Contract No.DE-SC0012704 and Award No. DE-SC0021969, respectively. C.S. acknowledges support from a DOE Office of Science Early Career Award.

\clearpage
\appendix
\section{Supplementary material}

\paragraph{Skyrme-HFB calculations}

We used the MOCCa code of Ref.~\cite{Ryssens16} to solve the self-consistent 
Skyrme-HFB equations, representing the nucleonic single-particle wavefunctions 
on a three-dimensional coordinate grid, resulting in an easily-controlled 
numerical accuracy that is independent of the nuclear shape considered~\cite{Ryssens15b}. 
The numerical conditions were
identical to those employed in Ref.~\cite{Scamps21}: both $^{197}$Au and $^{238}$U 
were represented on a Cartesian mesh with $N_X = N_Y = N_Z = 36$  points in each
direction, spaced equidistantly with $dx=0.8$ fm. Since a complete single-particle
basis on this mesh would require monstrous amounts of memory, we iterated
for each nucleus only the $(N+Z+340)$ single-particle states with lowest 
single-particle energy.

The parametrizations in the BSkG- and SV-families were used as originally 
published. For all other parametrizations we modified the pairing terms of the 
EDF. In the case of the SLy-family, no such terms were part of the parameter
adjustment at all. The UNEDF-family did include such terms but relied on 
a numerical representation in terms of harmonic oscillator basis functions, 
resulting in different pairing properties that cannot be reproduced by our
coordinate-space representation. In these cases, we employed the following 
simple form to introduce `surface-peaked' pairing terms in the EDF:

\begin{align}
E_{\rm pair} &= \sum_{q=p,n} \frac{V_{q}}{4} \int d^3\bold{r}
\, \left[1 - \left(\frac{\rho_0(\bold{r})}{\rho_{\rm sat}}\right) \right]  
\tilde{\rho}_q^*(\bold{r}) \tilde{\rho}_q(\bold{r}) \, ,
\end{align}
where $\rho_{\rm sat} = 0.16 $ fm$^{-3}$ and $\rho_0(\bold{r})$ is the isoscalar
density. The definition of the pairing densities $\tilde{\rho}_{q}(\bold{r})$ is 
standard in the literature (see for example Ref.~\cite{Ryssens21}): we calculate
them with cutoffs at 5 MeV both above and below the Fermi energy as in Ref.~\cite{Krieger90}. $V_n$ and $V_p$ are 
parameters that characterize the overall strength of neutron and proton pairing, 
respectively. All SLy-family parametrizations we employ here have similar
effective mass and so we use  $V_n = V_p = -1250 $ MeV fm$^{-3}$ for all of
them, following Ref.~\cite{Rigollet99}. For the UNEDF-family, we adjusted the pairing 
strengths to roughly reproduce the experimental three-point mass-staggering for 
protons and neutrons for $^{238}$U. This resulted in values
of $(V_n, V_p)$ of $(-850,-1250)$, $(-920, -1250)$ and $(-950,-1350)$ MeV fm$^{-3}$
for UNEDF0, UNEDF1 and UNEDF2 respectively. None of the results we report
here depend strongly on the values of these parameters.

To save on computational resources, we restricted our simulations to nuclear
configurations invariant under three plane reflections. Imposing these
self-consistent spatial symmetries allowed us to limit the explicit 
numerical representation to only one-eight of all mesh points. We also assumed the 
conservation of time-reversal symmetry in nearly all calculations, allowing us
to reduce the computational effort by another factor of two. The sole exception
was the BSkG2 calculation for $^{197}$Au, where we accounted for the full 
effect of the odd neutron: the breaking of time-reversal symmetry through the 
self-consistent blocking procedure and all so-called `time-odd' terms of the 
EDF~\cite{Ryssens22}. The latter are not well-defined for the other 
parametrizations, such that we relied on the equal filling approximation to 
perform self-consistent blocking calculations for $^{197}$Au
without breaking time-reversal symmetry in all other cases~\cite{Martin08}.
Irrespective of symmetry choices, we used a strategy based on the gradient 
algorithm of Ref.~\cite{Bertsch11} to construct the blocked state with minimum total 
energy after convergence. 

\paragraph{Wood-Saxon fits and supplementary files}

We adjusted the parameters of the WS form (Eq.~\eqref{eq:WSdensity}) to
reproduce the values of the total density $\rho_0(\bold{r})$ at the mesh 
points in the EDF calculation. We limited outselves to five deformation 
parameters: $(\beta^{\rm WS}_{20}, \beta^{\rm WS}_{22}, \beta^{\rm WS}_{40}, 
\beta^{\rm WS}_{42}, \beta^{\rm WS}_{44})$, which led to good fits for both 
$^{238}$U and $^{197}$Au. Allowing for the polarisation of the surface 
diffuseness as in Ref.~\cite{Scamps13} does not meaningfully change the 
extracted deformation parameters but does allow for a better fit. We omitted
this possibility as these degrees of freedom have so far not been studied in
hydrodynamic simulations of heavy ion collisions. A more modest improvement 
of the fit for $^{238}$U can be achieved by including the $\ell = 6$ deformation 
parameters, but these do not impact the quadrupole and hexadecapole deformations
much for this nucleus.

A complete set of the multipole moments $\beta_{\ell m}$ and fitted WS deformation 
parameters $\beta^{\rm WS}_{\ell m}$ for all Skyrme parametrizations is included in the supplementary
files \texttt{Au197.dat} and \texttt{U238.dat}. The structure of these files
is clarified by Tab.~\ref{tab:parameters}, where we also include as examples 
the values obtained for both nuclei with the BSkG2 parametrization~\cite{Ryssens22}.
For convenience, we also report the quadrupole deformation in terms of its
total size $\beta_2$ and the triaxiality angle $\gamma$. These are linked 
to the $\beta_{20}$ and $\beta_{22}$ moments 
through~\cite{Scamps21}:
\begin{align}
\beta_{2} &= \sqrt{\beta_{20}^2 + 2 \beta_{22}^2} \, , \\
\gamma    &= \text{atan} \left( \sqrt{2}\beta_{22}/ \beta_{20}\right)\, .
\end{align}
Analogous conversion relations apply to the WS deformation
parameters, $\beta_2^{\rm WS}$ and $\gamma^{\rm WS}$. 
\begin{table}
\begin{tabular}{lr@{\quad}d{4.3}@{\quad}d{3.2}@{\quad}}
\hline\noalign{\smallskip}
\hline\noalign{\smallskip}
                      &  Column  & \multicolumn{1}{c}{$^{197}$Au} & \multicolumn{1}{c}{$^{238}$U}\\ 
\hline\noalign{\smallskip}
$R_d$ (fm)            &  1 &   6.620    & 7.068  \\
a (fm)                &  2 &   0.519    & 0.538  \\
\hline\noalign{\smallskip}
 $\beta_{20}$         &  3 &  +0.089    &+0.280  \\
 $\beta_{22}$         &  4 &  +0.065    & 0.000  \\
 $\beta_{2}$          &  5 &  +0.128    &+0.280  \\
 $\gamma$ (deg)       &  6 &   45.9     & 0       \\
 $\beta_{40}$         &  7 &  -0.017    &+0.153  \\  
 $\beta_{42}$         &  8 &  -0.011    & 0.000  \\ 
 $\beta_{44}$         &  9 &  -0.010    & 0.000  \\    
\hline\noalign{\smallskip}
 $\beta^{\rm WS}_{20}$   & 10 &  +0.098    &+0.247  \\
 $\beta^{\rm WS}_{22}$   & 11 &  +0.076    & 0.000  \\
 $\beta^{\rm WS}_{2}$    & 12 &   0.145    & 0.247  \\
 $\gamma^{\rm WS}$ (deg) & 13 &   47.6     & 0       \\
 $\beta^{\rm WS}_{40}$   & 14 &  -0.025    &+0.081  \\ 
 $\beta^{\rm WS}_{42}$   & 15 &  -0.018    & 0.000  \\ 
 $\beta^{\rm WS}_{44}$   & 16 &  -0.018    & 0.000  \\    
\hline\noalign{\smallskip}
\hline\noalign{\smallskip}
\end{tabular}
\caption{ Multipole moments and best-fit WS parameters 
$R_d, a$ and $\beta^{\rm WS}_{\ell m}$ for the one-body 
           densities of $^{197}$Au and $^{238}$U, as obtained with the BSkG2
           Skyrme parametrization. The corresponding column numbers in the 
           supplementary files \texttt{Au197.dat} and \texttt{U238.dat} are
           indicated.
}
\label{tab:parameters}
\end{table}

\end{document}